\documentclass[aps,prd,superscriptaddress,showpacs, preprint]{revtex4-1}
\usepackage{graphicx}
\usepackage{epstopdf}
\usepackage{amsmath}
\usepackage{amsfonts}
\usepackage{amssymb}
\usepackage{latexsym}
\usepackage{hyperref}
\usepackage{xcolor}
\usepackage[colorinlistoftodos]{todonotes}

\begin{document}

\title{Comparative aspects of spin-dependent interaction potentials for spin-1/2 and spin-1 matter fields}

\author{P.C. Malta}\email{malta@thphys.uni-heidelberg.de}
\affiliation{Institut f\"ur theoretische Physik, University of Heidelberg, Philosophenweg 16, 69120 Heidelberg, Germany}
\affiliation{Centro Brasileiro de Pesquisas F\'{i}sicas (CBPF), Rua Dr Xavier Sigaud 150, Urca, Rio de Janeiro, Brazil, CEP 22290-180}

\author{L.P.R. Ospedal}\email{leoopr@cbpf.br}
\affiliation{Centro Brasileiro de Pesquisas F\'{i}sicas (CBPF), Rua Dr Xavier Sigaud 150, Urca, Rio de Janeiro, Brazil, CEP 22290-180}

\author{K. Veiga}\email{kimveiga@cbpf.br}
\affiliation{Centro Brasileiro de Pesquisas F\'{i}sicas (CBPF), Rua Dr Xavier Sigaud 150, Urca, Rio de Janeiro, Brazil, CEP 22290-180}
\affiliation{Instituto Federal de Educa\c{c}\~{a}o, Ci\^{e}ncia e Tecnologia da Bahia - Campus Vit\'{o}ria da Conquista, Av. Amazonas 3150, Zabel\^{e}, Vit\'{o}ria da Conquista, Bahia, Brazil, CEP 45075-265} 

\author{J.A. Helay\"{e}l-Neto}\email{helayel@cbpf.br}
\affiliation{Centro Brasileiro de Pesquisas F\'{i}sicas (CBPF), Rua Dr Xavier Sigaud 150, Urca, Rio de Janeiro, Brazil, CEP 22290-180}


\begin{abstract}

This paper sets out to establish a comparative study between classes of spin- and 
velocity-dependent  potentials for spin-1/2 and spin-1 matter 
 currents/sources in the non-relativistic regime. Both (neutral massive) scalar and 
vector particles are considered to mediate the interactions between   (pseudo-)scalar sources or (pseudo-)vector  currents.  Though our discussion is more general, we contemplate specific cases in which our results may describe the electromagnetic interaction with a massive (Proca-type) photon exchanged between two spin-1/2 or two spin-1 carriers. We highlight the similarities and peculiarities of the potentials for the two different types of charged matter and also focus our attention  on the comparison between  the particular aspects of two different field representations for spin-1 matter particles. We believe that our results may contribute to a further discussion of the relation between charge, 
spin and extensibility  of elementary particles.

\end{abstract}

\maketitle


\section{Introduction}

\indent

Back to 1950, the paper by Matthews and Salam \cite{M_S} and the subsequent works by 
Salam in 1952 \cite{S1} clarified the quantum field-theoretical approach to the electrodynamics 
of (massive) charged scalar particles. Ever since, the problem of extending this 
investigation to include the case of (massive) charged vector particles became mandatory 
in view of the theoretical evidence for the role of charged and neutral spin-1 particles 
that should couple to the charged and neutral currents.

From 1960, Komar and Salam \cite{K_S}, Salam \cite{S2,S3}, Lee and Yang \cite{L_Y} and Delbourgo 
and Salam \cite{D_S}, gave a highly remarkable push for the understanding and construction 
of a fundamental theory for the microscopic interaction between charged vector bosons and 
photons. Further works by Tzou \cite{T}, Aronson \cite{A} and Velo and Zwanziger \cite{V_Z} summed 
up efforts to the previous papers and the final conclusion was that a consistent unitary 
and renormalizable quantum field-theoretical model would be possible only in a 
non-Abelian scenario with a Higgs sector that spontaneously breaks the gauge symmetry to 
give mass to the vector bosons, without spoiling the unitarity bounds for the cross section 
of the scattering of longitudinally polarized charged vector bosons \cite{G}.

Here, we adopt this framework: the electrodynamics of (massive) charged vector bosons 
with a non-minimal dipole-type term coupling  to the gauge field. In the case of the massless photon, this ensures a $g = 2-$value for the g-ratio of spin-1 particles. We should however point out that this dipole-type 
interaction is non-minimal from the viewpoint of an Abelian symmetry; if we  take 
into account the local $SU(2)$ symmetry that backs the plus- and minus-charged vector 
bosons, the dipole coupling in the action is actually a minimal (non-Abelian, $SU(2)$) 
interaction term after spontaneous symmetry breaking has taken place.

In the present paper, our main effort  consists in pursuing a detailed investigation of the 
semi-classical aspects of the charge and spin interactions for massive charged matter of a 
vector nature.  Our central purpose is to compare the features and specific profiles of 
the influence of the spin of the charge carriers on 
interaction potentials (electromagnetic or a more general $U(1)$ interaction) between  two
different categories of  sources/currents:  fermionic and  spin-1 bosonic. We shall present the conclusions of our comparative procedure at the end of our calculations.

At this point, we highlight that the literature in the topic contemplates a great deal of articles 
discussing the structure of the electromagnetic current and electrodynamical aspects of 
spin-1 charged  matter \cite{Nieves}-\cite{Bjerrum_Bohr}.  We pursue an investigation of an issue not considered in 
 connection with spin-1 charged matter: the spin- and velocity-dependence of the 
interaction potential  associated with (pseudo-)scalar sources and pseudo-vector  currents that interact by exchanging scalar and vector mediators, respectively. For the spin-1, these specific cases  have not been addressed to in the literature. These extra sources/currents may not necessarily be associated to the electromagnetic interaction in that they do not follow from the $U(1)$ symmetry of the electromagnetism. We may  be describing a new force between these extra sources/current whose origin could be traced back to some more fundamental physics. The case of the usual vector current is reassessed here and our results match with the ones in the literature. It is worthy mentioning that fermionic sources/currents can display a wide range of spin- and velocity-dependent interparticle potentials \cite{dobrescu}\cite{moody}\cite{Grupo}, and several of them have been reconsidered in this article.

The electrodynamics of ordinary fermionic matter is very well understood, from the macroscopic to the quantum level. Scalar  and vector bosonic  charged matter, on the other hand, experience a richer variety of interactions when coupled to the electromagnetic field. So far, most of the theoretical and experimental literature dealing with macroscopic interactions consider only spin-1/2 matter, i.e., fermionic  sources/currents \cite{adelberger}-\cite{brown}. This preference is naturally due to the fact that ordinary charges in matter are carried by electrons. Elementary spin-1 charged particles are difficult to observe, since the only known examples are the $W^{\pm}$ gauge bosons, whose mass is too large - and lifetime too short - to allow direct inspection (its full width is $\Gamma_{W} \sim 2 \, GeV$ \cite{PDG}). Though not stable enough to be directly handled, elementary spin-1 particles have, in principle,  their own electrodynamics and it is of theoretical interest to study how it deviates, or not, from its fermionic counterpart.  On the other hand, at the atomic and nuclear level, it would be a good motivation to have a spin- and velocity-dependent expression for the electromagnetic potential between ionised spin-1 (charge $= \pm 1$) atoms and charged spin-1 nuclei or hypernuclei \cite{Agnello}\cite{Yamamoto}.


We highlight here a particular feature of bosonic charge carriers as far as the 
electromagnetic interactions are concerned. From a purely macroscopic point of view, the 
Maxwell equations address the problem of determining field configurations from given 
charge and current densities and a number of duly specified boundary conditions. They do 
not take into account the microscopic nature of  these sources ($\rho$ 
and $\vec{j}$, respectively). If a microscopic description of charged matter in terms of 
classical fields is given (based on a local $U(1)$ symmetry), the particular aspects of 
the charge carriers become salient and London-type terms \cite{london} may arise. We shall be more specific about this point at the final Section of our paper, where we render more evident the peculiarities of the spin of the charge carriers in the electromagnetic and general Abelian interactions.

Our paper is outlined as follows: in Section II, we introduce the concept of potential and briefly discuss the kinematics involved, as well as present the notations and conventions employed. In Section III, we calculate the  sources/currents and potentials, discussing the similarities between spin-1/2 and spin-1 matter. We consider, in Section IV, another possible field representation for the spin-1 charged carrier, namely via a rank-2 tensor field, and we also work out the corresponding   source/current$-$source/current interaction potentials. We present our conclusions and perspectives in Section V. Natural units are adopted throughout, where $\hbar = c = 1$, and we assume $\epsilon^{0123} = +1$.


\section{Methodology} 

\indent

\begin{figure}[h!]
\centering
\includegraphics[width=0.5\textwidth]{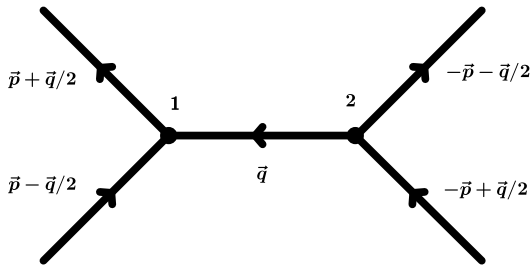}
\caption{\label{fig:overleaf} Momenta assignments for the tree-level interaction between  sources/currents 1 and 2.} \label{Fig1}
\end{figure}

Since we are interested in comparing the low-energy behavior of spin-1/2 and spin-1 as the interacting matter sector, it is convenient to work in the non-relativistic (NR) limit of the interactions. For simplicity we choose to work in the center of mass (CM) reference frame, in which particle 1 has initial and final 3-momenta given by $\vec{P} = \vec{p} - \vec{q}/2$ and $\vec{P'} = \vec{p} + \vec{q}/2$, respectively. Here, $\vec{p}$ is the average momentum of particle 1 before and after interaction, while $\vec{q}$ is the momentum transfer carried by the intermediate boson - see Fig.\eqref{Fig1}. 

The potential is calculated through the first Born approximation \cite{Mag}, i.e. the Fourier transform of the amplitude, $\mathcal{A}(\vec{q}, m\vec{v})$, with respect to the momentum transfer
\begin{equation}
V(r) = - \int\frac{d^{3}\vec{q}}{(2\pi)^{3}}\mathcal{A}(\vec{q}, \vec{p})e^{i\vec{q} \cdot \vec{r}}, \label{pot}
\end{equation}
where we explicitly assume that the amplitude contains terms dependent on the velocity of particle 1. It is clear from the NR approximations of the   sources/currents presented below that not only velocity-dependent terms will arise in the potentials, but also spin-dependent ones. This is due to our choice of keeping terms in the currents which go beyond the zeroth order, thus static, case \cite{Ac}. 

In what follows, we shall consider only elastic  interactions, which, together with energy conservation, imply $q^{0}=0$ and $\vec{p} \cdot \vec{q} = 0$, pointing out that $q$ is a space-like 4-vector: $q^{2} = - \vec{q}^{\, 2}$. Furthermore, in the amplitude we will keep only terms up to second order in $|\vec{p}|/m_{1,2}$. 


The Feynman rules for the tree-level diagram above are equivalent to taking $iJ_{1,2}$ as the interaction vertices, where $J_{1,2}$ are the matter  sources/currents associated with particles 1 and 2, respectively. The corresponding amplitude may then be written as
\begin{equation}
\mathcal{A} = i J_{1}\langle \text{prop.} \rangle J_{2}, \label{Amp}
\end{equation} 
where $\langle \text{prop.} \rangle$ is the momentum-space propagator of the intermediate boson with adequate Lorentz indices as the case may be. In this paper, we are interested in interactions mediated by massive neutral scalars (Klein-Gordon-type) and spin-1 vector particles (Proca-type), whose momentum-space propagators are:
\begin{equation} 
\langle \phi \phi \rangle = \frac{i}{q^{2} - m_{\phi}^{2}} \label{prop_s} 
\end{equation}
and
\begin{equation} 
\langle A_{\mu}A_{\nu} \rangle = \frac{-i}{q^{2} - m_{A}^{2}} \left( \eta_{\mu\nu} - \frac{q_\mu q_\nu}{m^2_A}\right) \, . \label{prop_Proca} 
\end{equation}

The sources/currents $J_1$ and $J_2$, as representatives of the respective vertices, must translate the different possible couplings to the gauge sector: scalar, pseudo-scalar, vector and pseudo-vector. The vector currents are obtained from the  Lagrangian through the Noether method, 
 but, for the bosonic case, we consider only first order in the electromagnetic coupling constant, $e$, since we wish to compute $\mathcal{A}$ to second order in $e$. We shall discuss this in more details in Subsection B.  The other sources/currents are built based on the principle that they should be bilinear in the field and its complex conjugate and reflect the desired symmetry.


Below, we recall some basic properties from spin-1/2 and spin-1 starting from the rest frame where the essential degrees of freedom become apparent, and then Lorentz-boosting it to the LAB frame. This will be important when we calculate the NR limit of each   source/current and then apply them to extract the interparticle potential. Here, we choose to work the spin-1 in its vector field representation, but, in Section IV, we shall discuss it in its less usual tensor field representation \cite{Delgado}\cite{Delgado_2}\cite{Napsuciale}.


\subsection{Spin-1/2}
\indent

For the sake of completeness, we present the well-known properties of Dirac fermions. 
The Dirac equation for the positive-energy spinors is, in momentum-space,
\begin{equation} 
\left[\gamma^\mu P_\mu - m\right] u(P) = 0 \label{Diraceq}
\end{equation}
with the corresponding equation for the spinor conjugate, $\bar{u} \equiv \, u^\dagger \gamma^0$, with the attribution $\bar{u} = \bar{u}(P')$.  By using  the gamma matrices in the Dirac representation, 
the positive-energy spinor can be explicitly written  in the NR limit as
\begin{equation} 
u(P) \simeq   N_f \left( \begin{array}{c} \xi \\ \frac{\vec{\sigma} \cdot 
\vec{P}}{2m} \, \xi\end{array} \right) ,\label{spinor}
\end{equation}
%
with $\xi$ are the basic spinors satisfying $\xi^{\dagger}_{r}\xi_{s} = \delta_{rs}$  and $N_f$ denotes a normalization constant. Manipulating eq.\eqref{Diraceq} and its conjugate, we obtain the  Gordon decomposition of the vector current, 
\begin{equation} 
\bar{u}  \gamma^\mu  u = \frac{p^\mu}{m} \, \bar{u} u + \frac{i}{m} \, q_\nu \, \bar{u} \, \Sigma^{\mu \nu} \, u  \label{GordonU}
\end{equation}
expressed in terms of the  more convenient variables $p \equiv (P+P')/2$ and $q \equiv P' - P$, where we defined the spin matrix as $\Sigma^{\mu \nu} \equiv  \frac{i}{4} \left[ \gamma^\mu , \gamma^\nu 
\right]$. Similar decompositions are also possible for other  types of bilinear forms: for example, using eq.\eqref{Diraceq} and its conjugate, we show that
\begin{equation}
q_\mu \, \bar{u} \gamma^\mu \gamma_5 u = 2 m \, \bar{u} \gamma_5 u \, .\label{dPV_PS}
\end{equation}
 
The Gordon decomposition of the vector current associated with the Noether theorem, eq.\eqref{GordonU}, yields a first term  proportional to the density of the field ($\sim |u|^{2}$), while the second one presents a coupling between momentum transfer, $q$, and the spin matrix, $\Sigma^{\mu \nu}$, of the respective representation. This connection will be responsible for the zeroth order magnetic dipole moment interaction of the lepton with gyromagnetic ratio $g=2$ \cite{Ryder}. Interestingly enough, this simple structure found for the Gordon decomposition for the spin-1/2 will have a perfect analog in the spin-1 case, if we correctly include a non-minimal coupling - see Subsection B.

 Our interparticle potentials shall be derived from the (effective) Lagrangian describing the interactions between the exchanged (scalar or vector) particles and the sources/currents of the fermionic particle. It reads as follows below:
\begin{equation}
\mathcal{L}_{int}^{s=1/2} = \bar{\psi}\left[ \left( g_{S} + g_{PS}i\gamma_5 \right)\phi + \left( g_{V}\gamma^{\mu} + g_{PV}\gamma^{\mu}\gamma_5 \right)A_{\mu}  \right]\psi, \label{Lag_g_ferm}
\end{equation}
where the different couplings are all dimensionless. The scalar (S) and pseudo-scalar (PS) sources are not conserved and, since we are dealing with a massive Dirac field, we also have a non-conserved pseudo-vector (PV) current (see eq.\eqref{dPV_PS}).  As pointed out by Boulware \cite{Boulware}, the Proca-type field, whenever coupled to a conserved current, leads to a renormalizable model. For this reason, we shall consider the coupling between the $PV-$ current and Proca-type field as an effective interaction. The UV divergence does not harm our purposes here, since we are interested in the low-energy regime described by the interparticle potential. Actually, our external (on-shell) sources/currents are tailored in the NR regime, meaning that we are working much below the scale where UV divergences show up. Henceforth we shall denote the coupling constant $g_{V} \equiv e$ whenever considering the electromagnetic interaction.



\subsection{Spin-1}

\indent

In the previous Subsection, we  have revised some  basic aspects concerning spin-1/2 particles. Here, we would like to  follow up with a similar discussion for massive charged spin-1 particles in order to comparatively study how differently (or not) spin-1/2 and spin-1   sources/currents interact in the NR limit. 

We start  off discussing the vector field representation for a massive charged spin-1 particle, whose dynamics can be obtained from the following  Lagrangian:
\begin{eqnarray}
\mathcal{L}_{vec.} = -\frac{1}{2} \, \mathcal{W}^{\ast}_{\mu\nu} \mathcal{W}^{\mu\nu} + m^{2} \, W^{\ast}_{\mu}W^{\mu} + ie(g-1)W_{\mu}^{\ast}W_{\nu}F^{\mu\nu} \label{LagW}
\end{eqnarray}
where $m$ is the mass and $\mathcal{W}_{\mu\nu}$ the  gauge-covariant field strength, given by $D_{\mu}W_{\nu} - D_{\nu}W_{\mu}$, with $D_{\mu} = \partial_{\mu} + ieA_{\mu}$. The last term in eq.\eqref{LagW} is a non-minimal coupling between the bosonic  matter fields and the field-strength of the interaction mediator and its importance lies in the fact that it asserts that $g=2$ in tree-level for the spin-1 particle \cite{Jackiw}\cite{Tel}\cite{Fidelman} (see also references therein).


The field $W^{\mu} = \left(W^{0}, \vec{W} \right)$ has, in principle, four degrees of freedom and these are reduced through the constraints introduced by the equations of motion, which read
\begin{equation}
D_{\mu} \mathcal{W}^{\mu\nu} + m^{2}W^{\nu} - ie(g-1)W_{\mu}F^{\mu\nu} = 0 \label{EqW}
\end{equation}
with  analogue ones for the complex conjugated field. Our goal is to characterize the   source/current$-$ source/current interaction and for this, at tree-level, the aforementioned free source/current are to be described by bilinears in the spin-1 matter fields, where the derivatives are the ordinary ones. In other words, since the asymptotic states are the particle states composed only of $W$ fields (we are now taking the free field equations), we have the subsidiary condition $\partial_{\mu}W^{\mu} = 0$. This constraint is better seen in momentum-space, where we have $\partial_{\mu}=-iP_{\mu}$, so that
\begin{equation}
W^{0} = \frac{1}{E}\vec{P}\cdot \vec{W}, \label{CW}
\end{equation}
where, in order to avoid a cumbersome notation, we have used the same symbol to denote the field in position- and momentum-space, as there is no risk of confusion. This equation relates different components of the vector field and shows that the time-component is proportional to the longitudinal projection of $\vec{W}$ and that $W^{0}$ is first-order in velocity, playing the role of a ``small'' component similar to the spin-1/2 case. It also allows us to write $W^{\mu}$ in its rest frame as $W^{\mu}_{rest} = N_{W}\left(0, \vec{\epsilon} \right) $, where $\vec{\epsilon}$ stands for the dimensionless polarization 3-vector  in the rest frame of the free particle and $N_{W}$ is a normalization constant. 

The next step is to bring the system from rest into motion (i.e., to the LAB frame) via an appropriate Lorentz transformation. In doing so, we obtain
\begin{equation}
W^{\mu}_{lab} = N_{W}\left[ \frac{1}{m}(\vec{P}\cdot\vec{\epsilon}) \,\, , \,\, \vec{\epsilon} + \frac{1}{m(E+m)}(\vec{P}\cdot \vec{\epsilon})\vec{P} \right] \label{Wlab}
\end{equation}
and with this, two comments are in order: 
\begin{itemize}
\item Eq.\eqref{CW} tells us that, in the rest frame (where $\vec{P}=0$), all the information about the vector field is contained in its spatial part. In this frame the only 3-vector available is the polarization, i.e., its spin, which again enforces the vector character of the field.

\item The normalization constant  is given by $N_W = 1/\sqrt{2m}$.

\end{itemize}

We wish to make some considerations on the vector currents, both global and local, associated to the Lagrangian \eqref{LagW}. At the global stage, i.e., prior to the gauging of the $U(1)$ symmetry, the Noether current  (in configuration-space) reads  
\begin{equation}  
J_{global}^\mu =  ie \left(  W^{\ast \mu \nu } W_\nu - W^{\mu \nu } W_\nu^\ast \right),
\end{equation}
where no covariant derivative is involved, so $W_{\mu \nu} \equiv \partial_\mu W_\nu - \partial_\nu W_\mu$. Upon the gauging of the symmetry, and by including the non-minimal coupling g-term, the current changes into
\begin{equation}  
J^\mu_V = ie \left(  \mathcal{W}^{\ast \mu \nu } W_\nu - \mathcal{W}^{\mu \nu } W^{\ast}_\nu \right) + ie(g-1) \partial_\nu \left( W^\nu W^{\ast \mu } - W^{\ast \nu} W^\mu \right) \, . \label{j_local}
\end{equation}

Now, an important remark concerning current conservation: to calculate the current-current potential at the tree-level approximation (one-boson exchange), as we are doing here, we actually only consider the current to the first order in $e$. This amounts to neglecting the terms in the gauge potential present in the current in eq.\eqref{j_local}. In other words: to calculate the potential to order $e^2$, we consider the current only up to order $e$. Therefore, to the desired order in the coupling constant, the  vector current is the globally conserved one, plus the non-minimal $g$-term. At this perturbative level, the current $J^\mu_V$ is given by the global and non-minimal terms, as displayed in the second line of eq.\eqref{Lag_g_bos} below: 
\begin{eqnarray}
\mathcal{L}^{s=1}_{int} & = & \left[ g_{S}W_{\mu}^{\ast}W^{\mu}  + g_{PS}W_{\mu\nu}^{\ast}\tilde{W}^{\mu\nu} \right]\phi + \nonumber \\
& + & ig_{V}\left[ W^{\ast \, \mu\nu}W_{\nu} - W^{\mu\nu}W^{\ast}_{\nu} + (g-1) \partial_{\nu}\left( W^\nu W^{\ast \mu} - W^{\ast \nu} W^\mu \right)   \right]A_{\mu} + \nonumber \\
& + & ig_{PV} \left[ \tilde{W}^{\ast \, \mu\nu}W_{\nu} - \tilde{W}^{\mu\nu}W^{\ast}_{\nu} \right]A_{\mu},  \label{Lag_g_bos}
\end{eqnarray} 
where the coupling constants have the following canonical mass dimensions: $[g_{S}] = - [g_{PS}] = 1$ and $[g_{V}] = [g_{PV}] = 0$, and the interactions between the  sources/currents and the mediating (scalar or vector) bosons are all displayed.
 
By using the equations of motion and the (free) subsidiary conditions, we may re-write the vector current in momentum space as
\begin{equation}
J_{V}^{\mu}(p,q) =  2ep^{\mu}W^{\ast}_{\nu}W^{\nu} + iegq_{\nu}W^{\ast \, \alpha} \left( \Sigma_{V}^{\mu\nu} \right)_{\alpha\beta}W^{\beta} \label{GordonW}
\end{equation}
in which $\left( \Sigma_{V}^{\mu\nu} \right)_{\alpha\beta} \equiv i\left( \delta^{\mu}_{\alpha}\delta^{\nu}_{\beta} - \delta^{\nu}_{\alpha}\delta^{\mu}_{\beta} \right)$ stands for the spin matrix in the $\left(\frac{1}{2},\frac{1}{2}\right)$-representation of the Lorentz group  and, as for the spin=1/2 case, we have set $g_V \equiv e$. In the equation above we left the g-factor explicit but in the NR limits we shall use its tree-level value, $g=2$. Eq.\eqref{GordonW} is nothing but the Gordon decomposition for the vector field representation of a massive and charged spin-1 and the similarity with the spin-1/2 case, eq.\eqref{GordonU}, is remarkable. This Gordon decomposition for spin-1 particles was found independently and agrees with the results of Ref.\cite{Delgado}. 


We denote the dual of $W_{\mu \nu}$, as given in the $PS-$ and $PV-$  couplings in eq.\eqref{Lag_g_bos}, by means of $\tilde{W}_{\mu \nu} \equiv \frac{1}{2} \, \epsilon_{\mu \nu \alpha \beta} W^{\alpha \beta}$. We also emphasize that the (global) $PV-$ bosonic current is conserved in a topological sense, i.e., without making use of either the equations of motion or the symmetries of the Lagrangian. As in the case of spin-1/2, the $S-$ and $PS-$  sources for spin-1 are  equally not conserved.

In the next Section, we discuss the  momentum-space sources/currents for spin-1/2 and spin-1  in the NR limit and calculate the interparticle potentials generated by the exchange of scalar and vector mediating bosons.


\section{Non-relativistic currents and potentials}

\indent

Following the indications above, we present below the NR limit of the   sources/currents, which may be extracted (in configuration-space) from Lagrangians \eqref{Lag_g_ferm} and \eqref{Lag_g_bos} as $ J\phi$ or $  J_{\mu}A^{\mu}$, as the case may be. In the following, the fields are already normalized with $N_f =1$ and $N_{W} = 1/\sqrt{2m}$. From now on, for the sake of clarity, we shall adopt  the variables $\left\{ p,q \right\}$ instead of $\left\{ P, P' \right\}$, so $u(P) = u(p-q/2)$ and $\bar{u}(P') = \bar{u}(p+q/2)$; similar definitions hold for $W_{\mu}(P)$ and $W^{\ast}_{\mu}(P')$. We point out that the matter   sources/currents developed in this Section are for particle 1,  i.e., the left vertex in Fig.\eqref{fig:overleaf}. The second source/current may be obtained by taking $\vec{q} \rightarrow -\vec{q}$, $\vec{p} \rightarrow -\vec{p}$ in the first one.

At this point it is convenient to reinforce the difference between polarization and spin of the particle. In the fermionic case, we shall denote the expectation value of its spin, $\sigma_i$, by contracting the  basic spinors $\xi$ with the Pauli matrices, namely, $\langle \sigma_i \rangle \equiv \xi^{\dagger} \sigma_i \xi $. The same idea is applied to the bosonic case, where the spin  matrix of the particle is $(S_i)_{jk} = i \epsilon_{ijk}$,  with $\Sigma_{ij} = \epsilon_{ijk} S_k$, and its expectation value is given by $\langle S_i \rangle \equiv \vec{\epsilon^{\ast}_j} \, (S_i)_{jk} \, \vec{\epsilon}_k$.

A remark  about our notation: in the process of calculating the amplitude factors, the form $\xi^{\dagger}\xi \equiv \delta_{i}$ and $\vec{\epsilon}^{\, \ast}\cdot \vec{\epsilon} \equiv \delta_{i}$, with $i=1,2$ labeling the particle, will appear frequently. These indicate a possible spin-flip between initial and final states, and we shall leave the $\delta$'s explicit in the final expressions, despite the fact that, in general, low-energy  interactions will not induce a change in the spin orientation of the particles involved. 


\subsection{Scalar (Klein-Gordon type) exchange}

\indent


By  means of the relations presented in the previous sections, we can carry out the NR expansion of the $S-$ and $PS-$ of the spin-1  sources. For the $S-$source of spin-1/2 particles, we can use the Dirac spinor, eq.\eqref{spinor}, and its conjugate, to obtain 
\begin{eqnarray} 
J_{S (s=1/2)} & = & g_{S}^1 \biggl\{ \delta_1 \left[ 1 - \frac{1}{4m_1^{2}}\left( \vec{p}^{\, 2} - \frac{1}{4}\vec{q}^{\, 2}  \right)   \right]   -  \frac{i}{4m_1^{2}}\vec{q}\cdot \left( \vec{p} \times \langle \vec{\sigma} \rangle_1 \right) \biggr\}  \label{us}.
\end{eqnarray}

For the $S-$source of the spin-1 particles, we  obtain the expression below:
\begin{eqnarray} 
W_\mu^* W^\mu = \frac{g^1_{S}}{2m_1}\left\{ -\delta_1 + \frac{1}{2m_1^{2}} \left[ (\vec{p} \cdot \vec{\epsilon}_1)\left(\vec{q}\cdot \vec{\epsilon_1^{\, \ast}}\right) - \left( \vec{p} \cdot \vec{\epsilon_1^{\, \ast}} \right)\left(\vec{q}\cdot \vec{\epsilon}_1 \right) -\left(\vec{q}\cdot \vec{\epsilon}_1 \right)\left(\vec{q}\cdot \vec{\epsilon_1^{\, \ast}}\right) \right] \right\},
\end{eqnarray}
which can be written in a more enlightening way if we note that  the polarization and momenta 3-vectors satisfy $- i(\vec{q} \times \vec{p}) \cdot \langle \vec{S} \rangle = (\vec{p} \cdot \vec{\epsilon})\left(\vec{q}\cdot \vec{\epsilon}^{\, \ast}\right) - (\vec{p} \cdot \vec{\epsilon}^{\, \ast})\left(\vec{q}\cdot \vec{\epsilon}\right)$. This allows us to set the spin-1 scalar source as 
\begin{eqnarray} 
J_{S (s=1)} & = & -\frac{g_{S}^1}{2m_1}\left\{ \delta_1 + \frac{1}{2m_1^{2}} \left[ i(\vec{q} \times \vec{p}) \cdot \langle \vec{S} \rangle_1 + \left(\vec{q}\cdot \vec{\epsilon}_1 \right)\left(\vec{q}\cdot \vec{\epsilon_1^{\, \ast}} \right) \right] \right\} \label{WS}.
\end{eqnarray}


By proceeding analogously, we can find the pseudo-scalar  sources listed below
\begin{eqnarray}
J_{PS (s=1/2)} & = & -\frac{ig^1_{PS}}{2m_1} \, \vec{q}\cdot \langle \vec{\sigma} \rangle_1 \label{ups} \\
J_{PS (s=1)} & = & ig_{PS}^1 \, \vec{q} \cdot \langle \vec{S} \rangle_1 \label{ps1},
\end{eqnarray}
and here we note that both pseudo-scalar sources have the same functional form  characterized by the coupling between spin and momentum transfer. In the following, we shall present the potentials.


Let us calculate a particular case, namely that of $S-S$ interaction, for both $s=1$ and $s=1/2$ sources. We start off with the fermionic case. As  discussed in the Methodology, the general amplitude is given by eq.\eqref{Amp} and, in this case,
\begin{equation} \mathcal{A}_{S-S}  = \frac{i }{q^{2} - m_{\phi}^{2}}\, J_{1} \, J_{2} , 
\label{amp_SS} 
\end{equation}
where we have used the Klein-Gordon propagator, eq.\eqref{prop_s}, and the momentum assignments of Fig.\eqref{fig:overleaf}:
\begin{eqnarray}
J_{1} & \equiv & g_S^1 \, \bar{u}(p + q/2) \,  u(p - q/2), \\
J_{2} & \equiv & g_S^2 \, \bar{u}(- p - q/2) \, u(- p + q/2).
\end{eqnarray}

At this point, we use the fermionic scalar source presented in eq.\eqref{us}. This scalar  source remains unchanged  by taking $q \rightarrow -q$ and $p \rightarrow -p$, since it is quadratic in the momenta (it will only be necessary a change in labels: $1 \rightarrow 2$). If we keep  terms up to second order in the momenta,  the amplitude  becomes
\begin{eqnarray}
\mathcal{A}_{S-S}^{s=1/2} & \simeq &  
\frac{g_S^1  g_S^2 }{\vec{q}^{\,2} + m^2_\phi} \Biggl\{ 
\delta_1 \delta_2 \left[ 1 -  \frac{1}{4}\left(\vec{p}^{\,2} - \frac{\vec{q}^{\,2}}{4} 
\right) \, \left( \frac{1}{m_1^2} + \frac{1}{m_2^2} \right) \right] + \nonumber \\
& - & \frac{i}{4} \, \left( \vec{q} \times \vec{p} \right) \cdot \left[ \frac{\delta_1 
\langle \vec{\sigma} \rangle_2}{m_2^2} + \frac{\delta_2 \langle 
\vec{\sigma} \rangle_1}{m_1^2} \right] \Biggr\}
\end{eqnarray} 
and the Fourier transformation given in eq.\eqref{pot} yields the potential:
\begin{eqnarray}
V_{S-S}^{s=1/2} & = & - \frac{ g_S^1 g_S^2}{4 \pi}\frac{e^{-m_\phi r}}{r}  \Biggl\{ \delta_1 \delta_2 \left[ 1 - 
\frac{1}{4}\left( \vec{p}^{\,2} + \frac{m_\phi^2}{4} \right) 
\left( \frac{1}{m_1^2} + \frac{1}{m_2^2} \right) \right] + \nonumber \\
& + &  \frac{\left( 1 + m_\phi r \right)}{4r^2} \vec{L} \cdot \left( 
\frac{\delta_1 \langle \vec{\sigma} \rangle_2}{m_2^2} + \frac{\delta_2 \langle 
\vec{\sigma} \rangle_1}{m_1^2} \right) \Biggr\}.    \label{S_S_ferm} 
\end{eqnarray}  

As expected, the dominant term is the monopole-monopole given by the Yukawa 
interaction, $-\delta_1 \delta_2  \, e^{-m_\phi r} /(4\pi r)$. We also obtain second-order corrections to this term in the form of a monopole-dipole with spin-orbit couplings. 

The amplitude of the bosonic $S-S$ case presents the same structure as given in eq.\eqref{amp_SS}, but with the  sources below: 
\begin{eqnarray}
J_{1} & \equiv & g_S^1  \, W^\ast_\mu(p + q/2) \,  W^\mu (p - q/2) , \\
J_{2} & \equiv & g_S^2 \, W^\ast_\mu(- p - q/2) \, W^\mu(- p + q/2). 
\end{eqnarray}

We follow the same program, i.e., by using the NR limit of the bosonic scalar  source, eq.\eqref{WS}, we obtain
\begin{eqnarray}
 \mathcal{A}_{S-S}^{s=1} & \simeq & \frac{g_S^1  g_S^2}{4m_{1}m_{2}\left(\vec{q}^{\,2} + m^2_\phi \right)} \Biggl\{ \delta_1 \delta_2 + \left[  \frac{\delta_1}{2m_2^2} \left[ i \left( \vec{q} \times \vec{p} \right) \cdot \langle \vec{S} \rangle_2
+ \left( \vec{q} \cdot \vec{\epsilon_2^\ast} \right) \left( \vec{q} \cdot 
\vec{\epsilon}_2 \right)  \right] + 1 \leftrightarrow 2 \right] \Biggr\} \nonumber
\end{eqnarray}
and, performing the Fourier integral, we get
\begin{eqnarray}
V_{S-S}^{s=1} & = & - \frac{g_{S}^{1}g_{S}^{2}}{16\pi m_{1}m_{2}} 
\frac{e^{-m_{\phi}r}}{r} \biggl\{ \delta_{1}\delta_{2} 
- \frac{(1 + m_{\phi}r)}{2r^2} \vec{L} \cdot \left( \frac{\delta_1 \langle 
\vec{S} \rangle_2}{m_2^2} + \frac{\delta_2 \langle 
\vec{S} \rangle_1}{m_1^2} \right) + \nonumber \\
& + & \zeta \left[  \frac{\delta_{1}}{2m_{2}^{2}r^{2}} \left[ \delta_2 (1 + 
m_{\phi}r) - \left(\hat{r}\cdot \vec{\epsilon_{2}}\right)\left(\hat{r}\cdot \vec{\epsilon_{2}^\ast} \right) 
\left( 3 + 3 m_{\phi} r + m_{\phi}^2 r^2 \right)   \right] 
+ 1 \leftrightarrow 2 \right]  \biggl\}, \label{S_S_rep_vet}
\end{eqnarray}
where the parameter $\zeta$ in the last term may assume the values $\zeta = \pm 1$; in eq.\eqref{S_S_rep_vet} its value is $\zeta = 1$. However, in the particular situation described in Section IV (a 2-form description of the massive spin-1 particle), the potential \eqref{S_S_rep_vet} also appears, but with $\zeta = -1$.

Here we notice that this potential describes the same interactions as in the fermionic case, 
namely, the Yukawa  factor and a spin-orbit  term, but it also displays a ``polarization-polarization" 
interaction term. This  observation will be present in other situations in our comparison between the fermionic 
and bosonic potentials. Below, we quote the results  for the other bosonic potentials:
\begin{eqnarray}
V_{S-PS}^{s=1} & = & \frac{g_{S}^{1}g_{PS}^{2}\delta_{1}}{8\pi m_{1}} 
\frac{(1+m_{\phi}r)e^{-m_{\phi}r}}{r^{2}} \hat{r}\cdot \langle \vec{S} \rangle_{2} \\ \label{S_PS_rep_vet}
V_{PS-PS}^{s=1} &  = & \frac{g_{PS}^{1}g_{PS}^{2}}{4\pi }\frac{e^{-m_{\phi}r}}{r} 
\biggl\{ \left( m_{\phi}^{2} + \frac{3m_{\phi}}{r} + \frac{3}{r^{2}} \right) \left( 
\hat{r}\cdot \langle \vec{S} \rangle_{1} \right)\left( \hat{r}\cdot \langle \vec{S} \rangle_{2} 
\right) + \nonumber \\
& - & \left( \frac{m_{\phi}}{r} + \frac{1}{r^{2}} \right) \langle \vec{S} \rangle_{1} 
\cdot \langle \vec{S} \rangle_{2}   \biggr\}.   \label{PS_PS_rep_vet}
\end{eqnarray}

The fermionic $PS-PS$ and $S-PS$ potentials have the same functional form as the 
bosonic ones and, if we neglect  a contact term (that is, a $\delta^3(\vec{r})$ term), these results  are very similar
 to those obtained by Moody and Wilczek \cite{moody} in the context of fermionic  sources  exchanging axions. Indeed, they only differ  by a mass factor due the canonical mass dimension of our coupling constants.


\subsection{Massive vector (Proca type) exchange}

\indent

Firstly, let us focus on the bosonic case, more precisely, on the spin-1 vector current, $J_V^\mu$.  For the $\mu = 0$ component in eq.\eqref{GordonW}, we have, in  momentum-space
\begin{eqnarray}
J^{0}_{V(s=1)} & = & -e_1 \left[ \delta_1 + \delta_1 \left( \frac{\vec{p}^{\, 2}}{2m_1^2} + \frac{\vec{q}^{\, 2}}{8m_1^2} \right) \right] + \nonumber\\ 
&+& (g-1)\frac{e_1}{2m_1^2} \left[ (\vec{p} \cdot \vec{\epsilon_1^{\, \ast}})(\vec{q} \cdot \vec{\epsilon}_1) - (\vec{q} \cdot \vec{\epsilon_1^{\, \ast}})(\vec{p} \cdot \vec{\epsilon}_1) + (\vec{q} \cdot \vec{\epsilon_1^{\, \ast}})(\vec{q} \cdot \vec{\epsilon}_1) \right],
\end{eqnarray}
where the entire second line comes from the $g$-factor correction term. If $g=1$, that line vanishes and we would have an incomplete description of the electromagnetic interaction experienced by the vector boson, as  seen from eq.\eqref{LagW}. Rearranging the equation above and taking  the correct $g=2-$value, we obtain the  time-component of the spin-1 vector current in the form
\begin{equation}
J^{0}_{V(s=1)} = -e_1  \delta_1 \left( 1 +  \frac{\vec{p}^{\, 2}}{2m_1^{2}} + \frac{\vec{q}^{\, 2}}{8m_1^{2}} \right)  + 
\frac{e_1}{2m_1^2} \left[ i(\vec{q} \times \vec{p}) \cdot \langle \vec{S} \rangle_1 + (\vec{q} \cdot \vec{\epsilon_1^{\, \ast}})(\vec{q} \cdot \vec{\epsilon}_1) \right], \label{WV0}
\end{equation}
while, by taking $\mu=i$ in eq.\eqref{GordonW}, we get the spatial component as given by
\begin{equation}
J^{i}_{V(s=1)} = -e_1 \left[ \frac{\delta_1}{m_1} \, \vec{p}_{i} + \frac{i}{m_1} \, \epsilon_{ijk}\vec{q}_{j}\langle \vec{S}_{k} \rangle_{1} \right], \label{WVi}
\end{equation}
which is very similar with its spin-1/2 counterpart, as we shall see in the following.

The spin-1 pseudo-vector current is simpler: according to its expression in eq.\eqref{Lag_g_bos}, we can obtain the following NR limits for $\mu = 0$ and $\mu = i$ as:
\begin{eqnarray}
J^{0}_{PV(s=1)} &=& \frac{ig^1_{PV}}{2m_1}\vec{q} \cdot \langle \vec{S}  \rangle_1 \label{WPV0} \\
J^{i}_{PV(s=1)} &=& -\frac{g^1_{PV}}{2m_1^{2}} \biggl\{ \left[ \left(\vec{p}-\frac{1}{2}\vec{q}\right) \cdot \vec{\epsilon}_1 \right] \left(\vec{q} \times \vec{\epsilon_1^{\, \ast}}\right)_{i} - \left[ \left(\vec{p}+\frac{1}{2}\vec{q}\right) \cdot \vec{\epsilon_1^{\, \ast}} \right]\left(\vec{q} \times \vec{\epsilon}_1 \right)_{i} \biggr\}. \label{WPVi}
\end{eqnarray} 

We present below the corresponding fermionic currents with the purely vector cases also carrying the electromagnetic coupling constant:
\begin{eqnarray} 
J^{0}_{V(s=1/2)} &=& e_1 \biggl\{ \delta_1 \left[ 1 + \frac{1}{4m_1^{2}}\left( \vec{p}^{\, 2} - \frac{1}{4}\vec{q}^{\, 2}  \right)   \right]   +  \frac{i}{4m_1^{2}} \left( \vec{q} \times \vec{p} \right)\cdot \langle \vec{\sigma} \rangle_1  \biggr\}  \label{uv0} \\ 
& \, & \nonumber \\
J^{i}_{V(s=1/2)} &=& e_1 \left[ \frac{\delta_1}{m_1}\vec{p}_{i} - \frac{i}{2m_1} \epsilon_{ijk}\vec{q}_{j}\langle \vec{\sigma}_{k} \rangle_1 \right],  \label{uvi} \\ 
& \, & \nonumber \\
J^{0}_{PV(s=1/2)} &=& \frac{g^1_{PS}}{m_1}\vec{p}\cdot \langle \vec{\sigma} \rangle_1, \label{upv0} \\ 
& \, & \nonumber \\
J^{i}_{PV(s=1/2)} &=& g^1_{PV} \biggl\{  \left[ 1 - \frac{1}{4m_1^{2}}\left( \vec{p}^{\, 2} - \frac{1}{4}\vec{q}^{\, 2}  \right)   \right]\langle \vec{\sigma}_{i} \rangle_1 -  \frac{i\delta_1}{4m_1^{2}} \left( \vec{q} \times \vec{p} \right)_{i} + \nonumber \\
& + & \frac{1}{2m_1^{2}} \left[ \left( \vec{p}\cdot \langle \vec{\sigma} \rangle_1 \right)\vec{p}_{i} - \frac{1}{4}\left( \vec{q}\cdot \langle \vec{\sigma} \rangle_1 \right)\vec{q}_{i} \right]  \biggr\}. \label{upvi} 
\end{eqnarray}

Unlike the $V-$ and $PV-$ currents of the spin-1/2 case, the $V-$ and $PV-$ currents of the spin-1 field, as defined in Lagrangian \eqref{Lag_g_bos}  (marked by the $V$ and $PV$ coupling constants), always carry an explicit derivative.  At high energies, this extra momentum factor must dominate, but not in the NR limit. This extra derivative affects the way  these currents interact with the gauge sector, specially in the case of spin-1. Furthermore, it is interesting to emphasize other differences between the fermionic and bosonic currents above. By comparing the vector currents, we notice essentially an extra term in the bosonic case associated with the contribution of the polarization, $(\vec{q} \cdot \vec{\epsilon})(\vec{q} \cdot \vec{\epsilon^\ast})$. The $PV-$currents display more remarkable differences. For example, for the $\mu = 0$ component, eqs.\eqref{WPV0} and \eqref{upv0}, we observe that the spin of the fermion couples to the average momentum, $\vec{p}$, while for the bosons the spin couples to the momentum transfer, $\vec{q}$. For the $\mu = i$ components, we have many differences due to the spin terms (see eqs.\eqref{WPVi} and \eqref{upvi}). These special features will be responsible for the different behaviors of the bosonic and fermionic potentials. 

Let us now discuss the potentials  involving $V-$ and $PV-$ currents mediated by a Proca-type particle. As already done in the previous Subsection, we shall  exemplify the calculation through a particular  configuration, 
and then  quote the final results. We choose to calculate one of the simplest cases, the 
$V-PV$ interaction between $s=1$  currents, since this case contains all the necessary  elements to understand 
the other (lengthier) evaluations.

The amplitude is given by 
\begin{equation} 
\mathcal{A}_{V-PV}^{s=1} = i J_1^\mu \, \langle A_\mu A_\nu \rangle \, J_2^\nu =- \frac{1}{\vec{q}^{\,2} + m_A^2} \, J_1^\mu \, J_{2 \mu},
\end{equation}
where we  have used current conservation, $ q^\mu J_\mu = 0$,  to eliminate the longitudinal contribution of the Proca-type propagator, eq.\eqref{prop_Proca}. 
In this particular case $(s=1)$, both currents are conserved. We need  therefore to simplify the contraction $J_1^\mu \, J_{2 \mu}$  and, according to our 
assumptions, only the $J_1^0 \, J_{2 \, 0}$  piece will contribute in the NR limit; so,
\begin{equation} 
\mathcal{A}_{V-PV}^{s=1} \simeq  - i \, \frac{e_1 g_{PV}^2 \delta_1}{2 m_2} \, \frac{\vec{q} 
\cdot \langle \vec{S} \rangle_2 }{\vec{q}^{\,2} + m_A^2} 
\, ,\end{equation}
and we  finally calculate the Fourier integral to obtain
\begin{equation}
V_{V-PV}^{s=1}  =  -\frac{e_{1} g_{PV}^{2}\delta_{1}}{8\pi m_{2}} 
\frac{(1+m_{A}r)e^{-m_{A}r}}{r^{2}} \hat{r}\cdot \langle \vec{S} \rangle_{2} \label{V_PV_rep_vet} \, .
\end{equation}

In a similar  way, the case with $s=1/2$  currents leads to
\begin{equation} 
V_{V-PV}^{s=1/2} = - \frac{e_1 g^2_{PV}}{4\pi} \, \frac{e^{-m_A r}}{r} \biggl\{ \delta_{1} \vec{p} 
\cdot \langle 
\vec{\sigma} \rangle_2  \biggl[ \frac{1}{m_1} + \frac{1}{m_2} \biggl] 
+ \frac{(1+m_A r)}{2m_1 r} [\langle \vec{\sigma} \rangle_1 \times \langle 
\vec{\sigma} \rangle_2 ] \cdot \hat{r} \biggl\}   \label{V_PV_ferm} 
\end{equation}
and here the $PV-$ current is not conserved; however, the conservation of the $V-$ current ensures that $q_\mu q_\nu$ term of the Proca propagator drops out.

As we have already seen in the discussion of the currents, the $PV-$ currents of $s=1/2$ 
and $s=1$  exhibit remarkable differences and these explain why the potentials behave so differently. 
We anticipate that only this fermionic case yields the  interaction with the
$ \vec{p} \cdot \langle \vec{\sigma} \rangle$ dependence and the spin-spin contribution of the 
form $\langle \vec{\sigma} \rangle_1 \times \langle \vec{\sigma} \rangle_2 $.  In the 
bosonic case, we have the $\hat{r}\cdot \langle \vec{S} \rangle$ dependence, but 
this is not exclusive of the $V-PV$  interaction, it also appears in the $S-PS$ bosonic potential. The other potentials, $V-V$ and $PV-PV$, are listed below. First, the fermionic ones:
\begin{eqnarray}
V_{V-V}^{s=1/2} & = & \frac{e_1 e_2}{4\pi} \, \frac{e^{-m_A r}}{r} \biggl\{ \delta_1 \delta_2 \biggl[ 1 + 
\frac{\vec{p}^{\,2}}{m_1 m_2} + \left( \frac{1}{4m_1^2} + \frac{1}{4m_2^2} \right) \left( \vec{p}^{\,2} + 
\frac{m_A^2}{4} \right) \biggl] + \nonumber \\
& - & \frac{(1+m_A r)}{2 r^2} \vec{L} \cdot \biggl[ \left( \frac{\delta_1 \langle 
\vec{\sigma} \rangle_2}{2m_2^2} + \frac{\delta_2 \langle \vec{\sigma} \rangle_1}{2m_1^2} \right) + 
\left( \frac{\delta_1 \langle \vec{\sigma} \rangle_2}{m_1 m_2} + \frac{\delta_2 \langle 
\vec{\sigma} \rangle_1}{m_1 m_2} \right)\biggl] + \nonumber \\
& + & \frac{1}{4m_1 m_2 r^{2}} \langle \vec{\sigma} \rangle_1 \cdot \langle \vec{\sigma} 
\rangle_2 \left( 1 + m_{A}r + m_{A}^{2}r^{2} \right) + \nonumber \\
& - & \frac{1}{4m_1 m_2 r^{2}} \left( \langle \vec{\sigma} \rangle_1 \cdot \hat{r} 
\right) \left( \langle \vec{\sigma} 
\rangle_2 \cdot \hat{r} \right) \left(3 + 3m_{A}r + m_{A}^{2}r^{2} \right)\biggl\} \, ,\label{V_V_ferm} 
\end{eqnarray}
where we can read the electromagnetic potential by setting $m_A = 0$; here $e_1$ and $e_2$ stand for the respective (electric) charges of the particles.

For the $PV-PV$ case we obtained
\begin{eqnarray}
V_{PV-PV}^{s=1/2} & = & -\frac{g^1_{PV} g^2_{PV} }{4\pi} \, \frac{e^{-m_A r}}{r} \biggl\{ \langle 
\vec{\sigma} \rangle_1 \cdot \langle \vec{\sigma} \rangle_2 \left[ 1 - \frac{1}{4}\left( \vec{p}^{\,2} 
+ \frac{m_A^2}{4} + \frac{m_A}{2r} + \frac{1}{2r^2} \right) \left( \frac{1}{m_1^2} + 
\frac{1}{m_2^2} \right) \right] + \nonumber \\
& + &  \left( \langle \vec{\sigma} \rangle_1 \cdot \vec{p} \right) \left( \langle 
\vec{\sigma} \rangle_2 \cdot \vec{p} \right) \left( \frac{1}{m_1 m_2} + \frac{1}{2m_1^2} + \frac{1}{2m_2^2} 
\right) + \nonumber \\
& + & \frac{\left(1+m_A r \right)}{4 r^2} \vec{L} \cdot \left( \frac{\delta_2 \langle \vec{\sigma} 
\rangle_1}{m_2^2} + \frac{\delta_1 \langle \vec{\sigma} \rangle_2}{m_1^2} \right) + \nonumber \\
& + & \frac{1}{8r^2} \left( \langle \vec{\sigma} \rangle_1 \cdot \hat{r} \right) 
\left( \langle \vec{\sigma} \rangle_2 \cdot \hat{r} \right) \left( \frac{1}{m_1^2} + \frac{1}{m_2^2} \right) 
\left( 3 + 3m_{A}r + m_{A}^{2}r^{2} \right)\biggl\} + V^{s=1/2}_{PV-PV-LONG} \, , \label{PV_PV_ferm}
\end{eqnarray}
where
\begin{eqnarray}  V_{PV-PV-LONG}^{s=1/2}  & \equiv &  \frac{g_{PV}^{1}g_{PV}^{2}}{4 \pi } \frac{e^{-m_{A}r}}{r} 
\biggl\{ \left( \frac{3}{m_A^2 r^2} + \frac{3}{m_{A}r} + 1 \right) \left( \hat{r}\cdot \langle 
\vec{\sigma} \rangle_{1} \right)\left( \hat{r}\cdot \langle \vec{\sigma} \rangle_{2} \right) + \nonumber \\
& - & \left( \frac{1}{m_A^2 r^2} + \frac{1}{m_{A} r}  \right) \langle \vec{\sigma} 
\rangle_{1} \cdot \langle \vec{\sigma} \rangle_{2} \biggr\}.    \label{LONG}
\end{eqnarray}  

The fermionic $PV-PV$ potential presents some peculiar aspects as an inheritance of the non-conserved $PV-$ currents. First of all, in contrast to other cases, we had to take into account the longitudinal part of the Proca propagator in our calculations. By doing that, we  obtain the contribution we have denoted by $V_{PV-PV-LONG}^{s=1/2}$ in eq.\eqref{LONG}. For this reason, this potential is not well defined in the massless limit $(m_A \rightarrow 0)$ of the exchanged particle and the excitation of the unphysical  spin-0 component of the massive vector boson ($\partial_\mu A^\mu$ does not decouple any longer) jeopardises the unitarity of the model under study.

In order to give a consistent interpretation and circumvent the  unitarity problem, we 
propose the following physical scenario: the exchanged boson is  relatively heavy $( m_1, m_2 > m_A \gg 
|\vec{q}|)$ and we are bound to work at energy and momentum-transfer levels much below 
$m_A$, so that we are far under the threshold for the excitation of the unphysical  longitudinal mode of
 $A_\mu$  as justified by the hierarchy of masses stated above. We highlight two 
aspects concerning the potential specified by eqs. \eqref{PV_PV_ferm} and \eqref{LONG}:

\begin{itemize}

\item We are working in the low momentum-transfer limit and, according to the uncertainty principle, 
one has  the interaction range approximated by $r \sim 1/|\vec{q}|$.  With this and the above hierarchy, we have $ r> \frac{1}{m_A} > \frac{1}{m_1} , \frac{1}{m_2} $ , 
which ensures that we are considering the potential for distances larger than the Compton wavelengths 
of the $s=1$ particles that act as external currents. This is consistent with the fact that the potential is a meaningful quantity for macroscopically large distances.  

\item Though our non-conserved  pseudo-vector current  couples to a massive vector boson, 
the fact that we are working in a low-energy regime  (NR approximation, 
having in mind  heavy $m_1$ and $m_2$) keeps us away from the danger of non-unitarity, which shows up in the 
high-energy domain $(E \gg m_A)$, so we conclude that the contribution described by eq.\eqref{LONG} is physically meaningful.
\end{itemize}

The spin-1 $V-V$ potential has the form:
\begin{eqnarray}
V_{V-V}^{s=1} & = & \frac{e_1 e_2}{4\pi} \frac{e^{-m_{A}r}}{r} \Biggl\{ 
\delta_{1}\delta_{2} \left[ 1+\frac{\vec{p}^{\,2}}{m_{1}m_{2}} + \left( \frac{1}{m_1^2} + 
\frac{1}{m_2^2} \right) \left( \frac{\vec{p}^{\,2}}{2} - 
\frac{m_{A}^2}{8}\right) \right] + \nonumber \\
& + &  \frac{(1+m_{A}r)}{r^{2}}\vec{L} \, \cdot \left[ \frac{1}{ m_1 
m_2} \left( \delta_{1} \langle \vec{S}\rangle_{2} + \delta_{2}\langle \vec{S}\rangle_{1} \right)
 +  \, \left( \frac{\delta_1}{2m_2^2} \langle \vec{S} \rangle_2 + 
\frac{\delta_2}{2m_1^2} \langle \vec{S} \rangle_1 \right) \right] + \nonumber \\
& + &  \frac{1}{m_{1}m_{2}r^2} \left[ \langle \vec{S}\rangle_1 \cdot \langle 
\vec{S} \rangle_2 \, \left( 1 + m_{A}r + m_{A}^{2}r^{2} \right) - \left(\hat{r} \cdot \langle \vec{S} \rangle_{1}\right) 
\left(\hat{r} \cdot \langle \vec{S} \rangle_{2}\right) \, \left( 3 + 3m_{A}r + m_{A}^{2}r^{2} \right) \right] + \nonumber \\
& - & \zeta \left[ \frac{\delta_1}{2m_2^2 r^2} \left( \delta_2 \left(1 + m_A r\right) - \left(  3 + 3m_{A}r + m_{A}^{2}r^{2}  \right) 
\, \left( \hat{r} \cdot \vec{\epsilon}_2 \right) \left( \hat{r} \cdot 
\vec{\epsilon^\ast_2} \right) \right) + \, 1 \leftrightarrow 2 \, \right] \Biggr\}, \label{V_V_rep_vet}  \end{eqnarray} 
where, as appears in eq.\eqref{S_S_rep_vet} for $V_{S-S}^{s=1}$, we have introduced the parameter $\zeta$. Again, its value in the vector field representation is $\zeta = +1$, whereas $\zeta =-1$ in the case of the tensor field parametrization. We shall be more specific about this point in the next section. The electromagnetic potential for the spin-1 charge carriers correspond to the choice $m_A=0$, with $e_1$ and $e_2$ denoting the electric charges of the particles.  In this case one clearly sees that the electromagnetic Coulomb interaction dominates at leading order for both spin-1/2 and spin-1  currents and, at this level, their respective potentials are indistinguishable. 


Finally, the bosonic $PV-PV$ potential 
\begin{eqnarray}  V_{PV-PV}^{s=1}  & = & -\frac{g_{PV}^{1}g_{PV}^{2}}{16\pi m_{1}m_{2}} \frac{e^{-m_{A}r}}{r^{3}} 
\biggl\{ \left( 3 + 3m_{A}r + m_{A}^{2}r^{2} \right) \left( \hat{r}\cdot \langle 
\vec{S} \rangle_{1} \right)\left( \hat{r}\cdot \langle \vec{S} \rangle_{2} \right) + \nonumber \\
& - & \left( 1 + m_{A}r  \right) \langle \vec{S} 
\rangle_{1} \cdot \langle \vec{S} \rangle_{2} \biggr\}    \label{PVPV1V}
\end{eqnarray}      
and here we recall that the $PV-$ current for $s=1$ is conserved, contrary to what happens with $s=1/2$. Therefore, this potential does not present the problems of its fermionic counterpart.

By comparing the results, we notice very similar spin-dependence in both $V-V$ 
potentials, such as spin-orbit and spin-spin interactions. However, the bosonic case has  additional terms with 
monopole-monopole and the polarization-polarization interactions (see the last term in eq.\eqref{V_V_rep_vet}).  From the fermionic $PV-PV$  potential, we observe an exclusive  spin-spin interaction with a $\left( \langle  \vec{\sigma} \rangle_1 \cdot \vec{p} \, \right) \left( \langle \vec{\sigma} \rangle_2 \cdot \vec{p} \, \right)-$dependence.  The bosonic case does not present these interactions and it reveals the same functional spin-dependence of the $PS-PS$ potential, eq.\eqref{PS_PS_rep_vet}.

To conclude this Section, in possession of the potentials we have calculated for spin-1 charged  currents/sources, we point out the possibility to apply our results to the study of the recently discovered heavy hyperhydrogen \cite{Agnello} and hyperhelium \cite{Yamamoto}. To study the details of the spectroscopy of the excited spin-1 states of $\Lambda-$hypernuclei, a careful analysis of the spin dependence of the potentials is needed (e.g. see Ref. \cite{Hashimoto}) and, in the particular case of the spin-1 states of the hypernuclei, such an analysis was actually lacking in the literature. Though we are not working out these applications in the present paper - it is out of our scope here - we endeavour to connect our results to the study of the excited spin-1/2 and spin-1 states of heavy hypernuclei. We shall report on that elsewhere.

\newpage 
\section{Tensor Field Representation}

\indent

We have discussed above the spin-1 particle in terms of the familiar vector field representation, but this is not the only possibility \cite{Delgado}\cite{Napsuciale}. On-shell, a massive charged spin-1 particle may well be described by a complex 2-form gauge field as given below:
\begin{eqnarray}
\mathcal{L}_{tens.} = -\frac{1}{6}\mathcal{G}^{*\mu\nu\kappa}\mathcal{G}_{\mu\nu\kappa} + \frac{1}{2} \, m^2B^{*\nu\kappa} B_{\nu\kappa} + ie(g-1){F_\nu}^\mu B_{\mu\kappa} B^{*\nu\kappa}, \label{LagB}
\end{eqnarray}
with $\mathcal{G}_{\mu\nu\kappa} = D_{\mu}B_{\nu\kappa} + D_{\kappa}B_{\mu\nu} + D_{\nu}B_{\kappa\mu}$ being the associated gauge-covariant 3-form field-strength. The equations of motion are found to be 
\begin{eqnarray}
D^\mu \mathcal{G}_{\mu\nu\kappa} + m^2 B_{\nu\kappa} + i e(g-1)B_{\mu\kappa} {F_\nu}^\mu + i e(g-1)B_{\nu\mu}{F_\kappa}^\mu = 0. \label{EqB}
\end{eqnarray}

Following similar arguments as in the Methodology, in the rest frame, we have
\begin{eqnarray}
B^{0i}_{rest} & = & 0 \\
B^{ij}_{rest} & = & \epsilon_{ijk} \vec{\epsilon}_{k},
\end{eqnarray}
where $\vec{\epsilon}$ is the polarization 3-vector in the rest frame of the particle. By suitably applying a Lorentz boost we attain the final form for $B^{\mu\nu}$ in the LAB system:
\begin{eqnarray}
B^{0i} & = & \frac{N_{B}}{m}\epsilon_{ijk}\vec{\epsilon}_{j}\vec{P}_{k} \\ \label{B0i}
B^{ij} & = & N_{B}\left\{ \epsilon_{ijk}\vec{\epsilon}_{k} + \frac{1}{m(E+m)} \left[ \vec{P}_{i}(\vec{\epsilon} \times \vec{P})_{j} - \vec{P}_{j}(\vec{\epsilon} \times \vec{P})_{i}  \right]\right\}, \label{Bij}
\end{eqnarray}
where the second term  in eq.\eqref{Bij} and the spatial component of eq.\eqref{Wlab} can be shown to be  analogous: it suffices to consider a specific direction and use the vector identity $\vec{a}\times ( \vec{b} \times \vec{c} ) = ( \vec{a}\cdot \vec{c} )\vec{b} - ( \vec{a}\cdot \vec{b} )\vec{c}$. The normalization  is given by  $N_{B} = 1/(2\sqrt{m})$.

 To study the matter sources and currents for this particular representation, we state below the interaction Lagrangian that contemplates all the couplings we are interested in:
\begin{eqnarray}
\mathcal{L}^{s=1}_{int} & = & \left[ g_{S}B_{\mu\nu}^{\ast}B^{\mu\nu}  + g_{PS}B_{\mu\nu}^{\ast}\tilde{B}^{\mu\nu} \right]\phi + \nonumber \\
& + & ig_{V}\left[ G^{\ast \, \mu\nu\kappa}B_{\nu\kappa} - G^{\mu\nu\kappa}B^{\ast}_{\nu\kappa} + (g-1) \partial^{\nu}\left(B_{\nu\alpha}B^{\ast \, \mu\alpha} - B^{\ast}_{\nu\alpha} B^{\mu\alpha} \right)   \right]A_{\mu} + \nonumber \\
& + & ig_{PV} \left[ G^{\ast \mu \nu \kappa} \tilde{B}_{\nu \kappa} - G^{ \mu \nu \kappa} \tilde{B}_{ \nu \kappa}^{\ast} \right]A_{\mu}, \label{Lag_g_bos_tens}
\end{eqnarray} 
 where the coupling constants have the following canonical mass dimensions: $[g_{S}] = [g_{PS}] = 1$ and $[g_{V}] = [g_{PV}] = 0$.  The 3-form field-strength, $G_{\mu \nu \kappa}$, does not include the gauge-covariant derivative. As previously discussed in connection with the gauge current in the vector field representation, the  vector current we present below is the one to order $e$ upon use of the subsidiary condition with the ordinary derivative, that is, the free subsidiary condition, which is compatible with our procedure to get the potential to order $e^2$. 

Following this prescription, we obtained the same result as Delgado-Acosta {\it et al} \cite{Delgado} in terms of the $\left\{ p,q \right\}$ momentum variables, namely, 
\begin{equation}
\mathcal{J}^{\mu}_{V}(p,q) = 2ep^{\mu}B_{\alpha\beta}^{\ast}B^{\alpha\beta} + iegq_{\sigma}B^{\ast \, \lambda\kappa} \left( \Sigma_{T}^{\mu\sigma} \right)_{\lambda\kappa,\alpha\beta}B^{\alpha\beta} , \label{GordonB}
\end{equation}
where
\begin{equation}
\left( \Sigma_{T}^{\kappa\sigma} \right)_{\mu\nu,\alpha\beta} = \frac{1}{2}\left[ \eta_{\mu\alpha} \left( \Sigma^{\kappa\sigma}_{V} \right)_{\nu\beta} - \eta_{\nu\alpha} \left( \Sigma^{\kappa\sigma}_{V} \right)_{\mu\beta} + \eta_{\nu\beta} \left( \Sigma^{\kappa\sigma}_{V} \right)_{\mu\alpha} - \eta_{\mu\beta} \left( \Sigma^{\kappa\sigma}_{V} \right)_{\nu\alpha}  \right]
\end{equation}
is the spin generator of the tensor field representation and,  from now on, the symbol $\mathcal{J}$ shall denote the momentum-space sources/currents in the tensor field representation. 


The NR  sources/currents in this representation read:
{\small \begin{eqnarray} 
\mathcal{J}_{S(s=1)} & = & \frac{g^1_{S}}{2m_1}\left\{ \delta_{1} + \frac{1}{2m_{1}^{2}} \left[ i(\vec{q} \times \vec{p}) \cdot \langle \vec{S} \rangle_1 + \delta_1 \vec{q}^{\, 2} - \left(\vec{q}\cdot \vec{\epsilon}_1 \right)\left(\vec{q}\cdot \vec{\epsilon_1^{\, \ast}}\right) \right] \right\} \label{BS} \\
\mathcal{J}_{PS(s=1)} & = & -\frac{ig^1_{PS}}{2m_{1}^{2}}\vec{q}\cdot \langle \vec{S} \rangle_{1} \label{ps2} \\
\mathcal{J}^{0}_{V(s=1)} & = & e_1 \left\{ \delta_{1}\left( 1 + \frac{\vec{p}^{\, 2}}{2m_1^2} + \frac{\vec{q}^{\, 2}}{8m_1^2} \right) - \frac{e_1}{2m^2_1} \left[ i(\vec{q} \times \vec{p}) \cdot \langle \vec{S} \rangle_1 + \delta_1 \vec{q}^{\, 2} - ( \vec{q} \cdot \vec{\epsilon}_1)\left( \vec{q} \cdot \vec{\epsilon_1^\ast} \right) \right] \right\} \label{BV} \\
\mathcal{J}^{i}_{V(s=1)} &=& e_1 \left[ \frac{\delta_{1}}{m_{1}}\vec{p}_{i} + \frac{i}{m_{1}} \epsilon_{ijk}\vec{q}_{j}\langle \vec{S}_{k} \rangle_{1} \right] \label{BVi} \\
\mathcal{J}^{0}_{PV(s=1)} &=& -\frac{ig^1_{PV}}{2m_{1}}\vec{q}\cdot \langle \vec{S} \rangle_{1} \\
\mathcal{J}^{i}_{PV(s=1)} &\sim & \mathcal{O}(v^{2}).  \label{BPV}
\end{eqnarray}}

The  sources/currents of the two different spin-1 representations are very similar in the spin sector. Basically, a global sign distinguishes one from the other in every case, but this sign does not interfere in the evaluations of the amplitude. The only exception is the spatial component of the pseudo-vector current, where we notice some differences in the functional form. Nevertheless, this  detail is irrelevant, because its contribution to the $PV-PV$ or $V-PV$ potentials leads to terms of order higher than two in the momenta, which we are currently ignoring in our calculations. For this reason, we refrain from giving the explicit (lengthy) form of $\mathcal{J}^{i}_{PV(s=1)}$ and only indicate its order.

The remarkable differences appear in the polarization sector of the scalar  source and time-component of the vector  current. We recall that, in the vector field representation, eqs.\eqref{WS} and \eqref{WV0}, we have obtained the following polarization-dependence:
\begin{equation}
(\vec{q} \cdot \vec{\epsilon}) (\vec{q} \cdot \vec{\epsilon^{\ast}}) = \vec{q}^{\,2} \left( \frac{\vec{q}_i \vec{q}_j}{\vec{q}^{\,2}} \right) \vec{\epsilon}_i \vec{\epsilon^{\ast}_j} \, ,
\label{pol_long} \end{equation}
while in the tensor case, see eqs.\eqref{BS} and \eqref{BV}, we have
\begin{equation}
\delta \vec{q}^{\,2} - (\vec{q} \cdot \vec{\epsilon}) (\vec{q} \cdot \vec{\epsilon^{\ast}}) = \vec{q}^{\,2} \left( \delta_{ij} -  \frac{\vec{q}_i \vec{q}_j}{\vec{q}^{\,2}} \right) \vec{\epsilon}_i \vec{\epsilon^{\ast}_j} \, .
\label{pol_trans} \end{equation}

According to these results, the two representations differ also due to opposite projections in the contribution coming from  the polarization; this could be associated with the particular representation of the spin-1 we are dealing with. Let us be more specific: in order to evaluate the NR limit of these particular  cases, we had to take into account the  ``weak" contributions  (those with momenta) of the fields, such as the components $W^0 \sim \vec{P} \cdot \vec{\epsilon}$, $B^{0i} \sim (\vec{P} \times \vec{\epsilon})_i$, see eqs.\eqref{Wlab} and \eqref{Bij}, respectively, and the analogues present in $\vec{W}$ and $B^{ij}$. These different couplings between momenta and polarization, $\vec{P} \cdot \vec{\epsilon}$ and $\vec{P} \times \vec{\epsilon}$, arise due to the Lorentz boost to the LAB frame and produce the observed longitudinal and transverse projections appearing in eqs.\eqref{pol_long} and \eqref{pol_trans} above.

 These results (eqs.\eqref{pol_long} and \eqref{pol_trans}) raise a question concerning the vector or axial character of the spin-1 particle. In the $W^\mu$-case, we are assuming a vector-like particle. In the tensor representation, let us take the components as $B^{0i} = - B_{0i} \equiv \vec{X}_i$ and $B^{ij} = B_{ij} \equiv \epsilon_{ijk} \vec{Y}_k $. On-shell, in the rest frame of the particle, all the degrees of freedom (d.f.) are carried by $\vec{Y} \, (\text{with} \, \vec{X} = \vec{0})$. Going over to the LAB-system, the tensor representation gives $\vec{X} \sim \vec{P} \times \vec{Y}$. So, in a frame-independent way, the on-shell d.f. are actually described by $\vec{Y}$. The form of the vector current is insensitive to the vector or axial behavior of $\vec{Y}$, as we may conclude by a close inspection of the  $\mathcal{J}^\mu$ current expressed in terms of $\vec{X}$ and $\vec{Y}$. If $\vec{Y}$ is vector-like, then $\vec{X}$ must be axial, and vice-versa. Therefore, we confirm that the minus sign difference highlighted above is always present as  a consequence of the choice of representation (vector or tensor), regardless of whether the particle is vector- or axial-like.

At this point, it is instructive to discuss why we do not observe functional differences in all  sources/currents, but only a global (numerical or mass) factor. For example, by comparing the spatial component of the vector currents, eqs.\eqref{WVi} and \eqref{BVi}, we notice that the first contribution is of order $\mathcal{O}(v)$. One can check that the next correction should be $\mathcal{O}(v^3)$, which we are neglecting in this work. Similar arguments apply to the other cases. Therefore, we  could only achieve possible different profiles in these currents if we consider the relativistic corrections to higher orders. 

By using the same methodology as in Section III it can be shown that the $V-PV$ 
and $PV-PV$ potentials are identical to the results obtained in the 
vector field representation, eqs.\eqref{V_PV_rep_vet} and \eqref{PVPV1V}, while
the $S-PS$ and $PS-PS$ only differ by global mass factors, $(1/2m_2^2)$ and $(1/4m_1^2m^2_2)$, respectively, due to the 
normalization and the different canonical mass dimension of the coupling constant $g_{PS}$ (see eqs.\eqref{Lag_g_bos} and \eqref{Lag_g_bos_tens}). The other two potentials, $S-S$ and $V-V$, have, respectively, the following profiles:
\begin{equation}
V^T_{S-S} = V_{S-S}^{s=1} \Big|_{\zeta = -1} + \frac{g_{S}^{1}g_{S}^{2}}{16\pi m_{1}m_{2}} 
\frac{e^{-m_{\phi}r}}{r} \left[ \delta_{1}\delta_{2} \, \frac{m_\phi^2}{2} \, \left( 
\frac{1}{m_1^2} + \frac{1}{m^2_2} \right) \right]
\label{S_S_tensor} 
\end{equation}
\begin{equation} 
V_{V-V}^T =  V_{V-V}^{s=1} \Big|_{\zeta = -1} + \frac{e_1 e_2}{4\pi} 
\frac{e^{-m_{A}r}}{r} \left[ \delta_1 \delta_2 \, \frac{m_A^2}{2} \left( \frac{1}{m_1^2} + \frac{1}{m^2_2}  
\right) \right].
\end{equation}

The difference between the two representations in both $S-S$ and $V-V$ cases appears as monopole-monopole terms and results in opposite behaviors in the polarization-polarization sector, since $\zeta = +1$ and $-1$ in the vector and tensor representation, respectively. Actually, these are consequences of what we have seen in the discussion above about the scalar {\bf source} and time-component of the vector  current.

In general, we cannot distinguish between the vector and tensor field representations for the $s=1$ by only considering the spin-dependent sector of the potentials. We  emphasize that the differences are suppressed because they only show up in order $\mathcal{O}(v^2)$ in the amplitude.
 This is so because we are here bound to consider only dipole contributions. If we extend our calculations to also include quadrupole effects, the differences in the $V-V$ potentials for the vector and tensor representation of the spin-1 might become evident, as we could expect from the studies reported in Ref. \cite{Delgado}. The consequences of the quadrupole moments of the two representations are interesting to be investigated and we intend to report on that in a forthcoming work \cite{Grupo_2}.


\section{Concluding remarks}

\indent

In this paper, we focused our attention on the special properties of the spin-spin 
dependence of the  source/current$-$source/current interaction potential in the non-relativistic limit for  spin-1 and spin-1/2 charged matter particles. Our results indicate some universalities between bosons and fermions when exchanging scalar and vector particles. Essentially, we observe very similar contributions in the spin-sector of the $S-S$, $S-PS$, $PS-PS$ and $V-V$ potentials. The main differences show up in connection with $PV-$ currents, namely, the $V-PV$ and $PV-PV$ potentials.

We highlight here a result presented in Section IV: the non-trivial consequences of choosing a particular field representation for the spin-1 particle. Even if the vector and tensor representations are equivalent  on-shell (and the particles in the  source/current are on-shell), the results for the $S-S$ and $V-V$ potentials do differ. The justification was given in the aforementioned Section and we claim that, by including higher-order terms in the relativistic corrections,  quadrupole moments appear and the differences become explicit.

Furthermore, we do not present a discussion on the bosonic and fermionic (pseudo-)tensor currents exchanging a 2-form gauge field. However, according to preliminary results of a work in progress, it is worthwhile to comment that the case of bosonic  sources/currents would yield interesting new interactions involving spin-polarization couplings, e.g. $\vec{\epsilon_1} \cdot \langle \vec{S} \rangle_2$.

To close this report, we  point out a question that might be addressed to with the help of our results on the comparison between the classes of potentials describing the interaction between spin-1/2 and spin-1 matter sources/currents. For the fermionic case, the electromagnetic potential does not couple to the particle density, as it happens instead for spin-0 and spin-1 charged matter.

If a charged spin-1 field, $W^\mu$, is  non-minimally coupled  (see eq.\eqref{LagW}) to the electromagnetic field, we have 
$$ \partial_\mu F^{\mu \nu} - 2 e^2 \left( W_\mu^\ast W^\mu \right) A^\nu
+ e^2 \left[ W^{\ast \nu } \left( W_\mu A^\mu \right)  +   W^{\nu} \left( W_\mu^\ast A^\mu \right) \right]  =  $$
\begin{equation}
 = J_{global}^\nu + ie \partial_\mu \left( W^\mu W^{\ast \nu } - W^{\ast \mu} W^\nu 
\right) \label{maxW}
\end{equation}
where $ J_{global}^\nu = -ie \left( W_\mu^\ast W^{\nu \mu} - W^{\ast \nu \mu } W_\mu \right)$ and $ W_{\nu \mu} \equiv \partial_\nu W_\mu - \partial_\mu W_\nu $.  The second term of the RHS contributing to $J^{\nu}$ stems from the non-minimal coupling from eq.\eqref{LagW} and is mandatory to ensure $g=2$ for the charged spin-1 boson.

From eq.\eqref{maxW}, two interesting properties of the electromagnetic  interactions of spin-1  matter fields  are made explicit: the London-type term that couples $A^\mu$ to the density of carriers and the new interaction 
between the photon and the charged boson through their polarization vectors. These special aspects are expected to influence the inter-particle potential only  if we take loop effects into account.

 However, the $e^2-$terms in eq.\eqref{maxW} introduce singularities associated to the point-like idealization of spin-1 charged matter particles. The 1962 paper by Dirac \cite{Dirac_62} and the recent works by Fabbri \cite{Fabbri} and Dain \cite{Dain} address the issue of extensibility in connection with charged particles.

In the present paper, we do not calculate classical field configurations generated by spin-1 
currents; instead, we build up interaction potentials by means of a semi-classical 
calculation, so we do not run into the complications yielded by the singularities mentioned above. It would be nevertheless worthwhile to analyse the details of the connection between particle extension, mass, charge and spin. The electrodynamics of bosonic carriers seems to suggest that the point-like idealization 
of charged particles is indeed very restrictive.
 
We hope to carry on with our studies in this direction and to make use of our results for the 
spin-1 potentials to get a deeper insight in the understanding of the electrodynamical 
properties of charged bosonic fields in the classical regime.


\begin{acknowledgments}
We would like to thank the National Council for Scientific and Technological Development of Brazil (CNPq), the Co-ordination for Qualification of Higher Education Personnel (CAPES), the German Service for Academic Exchange (DAAD) and the Funda\c{c}\~{a}o Carlos Chagas Filho de Amparo \`{a} Pesquisa do Estado do Rio de Janeiro (FAPERJ) for the invaluable financial support.  The authors also express their appreciation to the (anonymous) Referees for the pertinent and constructive criticisms and suggestions on the original manuscript.
\end{acknowledgments}


\end{document}


\title{Corrigendum to ``Comparative aspects of spin-dependent interaction potentials for spin-1/2 and spin-1 matter fields"}

\author{P.C. Malta}\email{malta@thphys.uni-heidelberg.de}
\affiliation{Institut f\"ur theoretische Physik, University of Heidelberg, Philosophenweg 16, 69120 Heidelberg, Germany}
\affiliation{Centro Brasileiro de Pesquisas F\'{i}sicas (CBPF), Rua Dr Xavier Sigaud 150, Urca, Rio de Janeiro, Brazil, CEP 22290-180}

\author{L.P.R. Ospedal}\email{leoopr@cbpf.br}
\affiliation{Centro Brasileiro de Pesquisas F\'{i}sicas (CBPF), Rua Dr Xavier Sigaud 150, Urca, Rio de Janeiro, Brazil, CEP 22290-180}

\author{K. Veiga}\email{kimveiga@cbpf.br}
\affiliation{Centro Brasileiro de Pesquisas F\'{i}sicas (CBPF), Rua Dr Xavier Sigaud 150, Urca, Rio de Janeiro, Brazil, CEP 22290-180}
\affiliation{Instituto Federal de Educa\c{c}\~{a}o, Ci\^{e}ncia e Tecnologia da Bahia - Campus Vit\'{o}ria da Conquista, Av. Amazonas 3150, Zabel\^{e}, Vit\'{o}ria da Conquista, Bahia, Brazil, CEP 45075-265} 

\author{J.A. Helay\"{e}l-Neto}\email{helayel@cbpf.br}
\affiliation{Centro Brasileiro de Pesquisas F\'{i}sicas (CBPF), Rua Dr Xavier Sigaud 150, Urca, Rio de Janeiro, Brazil, CEP 22290-180}


\begin{abstract}
In this short note we present punctual corrections to the paper ``Comparative aspects of spin-dependent interaction potentials for spin-1/2 and spin-1 matter fields", Adv. High Energy Phys. {\bf 2016}, (2016) 2531436 (arxiv:hep-th/1510.03291v4). Some of the interparticle potentials receive small corrections, mainly as monopole or spin-dependent contact terms. The main conclusions remain basically unaltered.
\end{abstract}

\maketitle


\section{Introduction} \label{sec_1}

\indent

In our work \cite{grupo} we calculated the interparticle potentials for sources with spin-1/2 and spin-1 (in the vector and tensor representations) in the non-relativistic (NR) regime. Our objective was to understand the role of the spin in these interactions, a task only possible if one goes beyond the static limit in the amplitudes.

The passage from the fully relativistic to the NR amplitude is accomplished via \cite{Maggiore}
\begin{equation}
\mathcal{M}_{\rm NR} = \displaystyle \prod_{i = 1,2} \left(2E_i\right)^{-1/2}  \displaystyle \prod_{j = 1',2'} \left(2E_j\right)^{-1/2} \mathcal{M} \label{M_NR}
\end{equation}
where, in \cite{grupo}, we used $\mathcal{M}_{\rm NR} \equiv \mathcal{A}$. Here $\mathcal{M}$ is the amplitude obtained from the corresponding Feynman diagram. The energy-dependent pre-factors are corrections to the NR normalization of states. These factors are fundamental to correctly obtain the NR interparticle potentials up to a given order -- this is specially true if we want to go up to $\mathcal{O}(|\vec{p}|^2/m^2)$, as is the objective in \cite{grupo}. There we erroneously approximated the $1/\sqrt{2E}$ factors by $1/\sqrt{2m}$, which was inconsistent with the approximation level of $\mathcal{O}(|\vec{ p}|^2/m^2)$ used for the rest of the amplitudes (and sources/currents). Incidentally, these factors were accounted for in the overall normalization of the wave functions, hence the normalizations stated in \cite{grupo}: $N_f = 1$, $N_W = 1/\sqrt{2m}$ and $N_B = 1/2\sqrt{m}$.

For the sake of completeness, we explicitly state the wave functions and their correct relativistic normalizations. The positive-energy solution for a spin-1/2 fermion (eq.(6) in \cite{grupo}) is
\begin{equation}
u(P) = \sqrt{E+m} \left( \begin{array}{c}
\xi \\\frac{\vec{\sigma}\cdot \vec{P}}{E+m} \xi  \end{array} \right) \label{spinor}
\end{equation}
with $\xi$ being the basic spinor normalized to $ u^\dagger(P) u(P) = 2E$. For the vector representation of a massive spin-1 boson (eq.(13) in \cite{grupo}), we have
\begin{equation}
W^\mu(P) = \left[\frac{1}{m} \vec{P}\cdot\vec{\epsilon} \,\, , \, \vec{\epsilon} + \frac{1}{m(E+m)} (\vec{P}\cdot\vec{\epsilon})\,\vec{P} \right], \label{W}
\end{equation}
whereas, for the tensor representation (eqs.(52) and (53) in \cite{grupo}):
\begin{eqnarray}
B^{0i} & = & \frac{1}{\sqrt{2}m}\epsilon_{ijk}\vec{\epsilon}_j \vec{P}_k \label{B0i} \\
B^{ij} & = & \frac{1}{\sqrt{2}} \Bigg\{  \epsilon_{ijk}\vec{\epsilon}_k + \frac{1}{m(E+m)} \left[ \vec{P}_i \left(  \vec{\epsilon} \times \vec{P} \right)_j - \vec{P}_j \left(  \vec{\epsilon} \times \vec{P} \right)_i  \right] \Bigg\}. \label{Bij}
\end{eqnarray}

In the two previous cases, $\vec{\epsilon}$ stands for the polarization 3-vector. Furthermore, in order to satisfy the spin algebra for $s=1$, we have to use $(S_i)_{jk} = -i \epsilon_{ijk}$, but in our work we erroneously considered the spin operator with the opposite sign, so one should take $S_i \rightarrow - S_i$ in our previous results.


\section{Correcting the potentials} \label{sec_corr}

\indent

With the observations above we are able to devise an effective correction factor for each of the three cases discussed in \cite{grupo}. Since there we assumed $N_f = 1$, we must include the correct normalization, $N_f = \sqrt{E+m}$, as well as the $1/\sqrt{E}$ factors from eq.\eqref{M_NR} -- similar considerations apply to the case of spin-1 fields. The ratio of these factors is approximately one up to $\mathcal{O}(|\vec{p}|/m_{1,2})$, but not up to $\mathcal{O}(|\vec{p}|^2/m_{1,2}^2)$, and this will induce extra momentum-dependent contributions to the amplitudes, therefore modifying some parts of the potentials.

Besides these corrections, we shall also include contact terms coming from Dirac delta contributions of the following Fourier transforms:
\begin{eqnarray}
\int \frac{d^3 \vec{q}}{(2\pi)^3} \frac{\vec{q}_i \vec{q}_j}{\vec{q}^{\,2} + m^2} \, e^{i \vec{q} \cdot \vec{r}} & = & \frac{\delta_{ij}}{3} \,\delta^{(3)}\left(\vec{r} \right) 
+ \frac{e^{-mr}}{4 \pi r^3} \,\left[ \delta_{ij} (1+mr) - \hat{r}_i \hat{r}_j (m^2 r^2 + 3mr + 3)  \right] \label{qiqj} \\
\int \frac{d^3 \vec{q}}{(2\pi)^3} \frac{\vec{q}^{\, 2}}{\vec{q}^{\,2} + m^2} \, e^{i \vec{q} \cdot \vec{r}} & = & \delta^{(3)}\left(\vec{r} \right) 
- \frac{m^2 e^{-mr}}{4 \pi r}, \label{q2}
\end{eqnarray}
which are the massive generalizations of the ``massless'' results from \cite{Adkins}.

In what follows we list the corrections -- indicated by $\Delta V$ -- that must be {\it added} to some of the potentials given in \cite{grupo}. The mass of the mediator, $m$, corresponds to either $m_{\phi}$ or $m_A$, as the case may be. We also highlight here that not all the potentials from \cite{grupo} are affected and in this note we only indicate those which are modified.




\subsection{Potentials (spin-1/2)} \label{sec_spin_12}

\indent

The correction factors for spin-1/2 sources usually do not affect monopole-dipole terms, as these originate from the lower, momentum-dependent, ``small'' component of the positive-energy spinor (cf. eq.\eqref{spinor}). Accordingly, the $S-PS$ and $V-PV$ potentials from \cite{grupo} are not affected.

The other potentials receive partial corrections, specially in the monopole-monopole and dipole-dipole terms. For the sake of convenience, we define the function
\begin{equation}
I(r) \equiv \frac{1}{4} \, \left( \frac{1}{m_1^2} +\frac{1}{m_2^2} \right) \, \left( \vec{p}^{\, 2} - \frac{m^2}{4} \right) \frac{e^{-mr}}{4 \pi r}
\end{equation}
so that the corrected potentials, eqs.(25), (30)\footnote{As mentioned below eq.(30) in \cite{grupo}, the PS-PS potential for $s = 1/2$ has the same functional form as the PS-PS potential for $s = 1$.}, (44) and (45) in \cite{grupo}, are respectively:

\begin{equation}
\Delta V_{S-S}^{s=1/2} = g_S^1 g_S^2 \, \delta_1 \delta_2 \, I(r)
\label{S-S_s=1/2} 
\end{equation}

\begin{equation}
\Delta V_{PS-PS}^{s=1/2} = - g_{PS}^1 g_{PS}^2 \, \frac{\langle \vec{\sigma} \rangle_1 \cdot  \langle \vec{\sigma} \rangle_2}{12\,  m_1 m_2} \, \delta^{(3)}\left(\vec{r} \right)
\label{PS-PS_s=1/2} 
\end{equation}

\begin{equation}
\Delta V_{V-V}^{s=1/2} = - e_1 e_2 \left\{ \, \delta_1 \delta_2 \, I(r) + \left[ \frac{\delta_1 \delta_2}{8} \left( \frac{1}{m_1^2} +\frac{1}{m_2^2} \right) + \frac{ \langle \vec{\sigma} \rangle_1 \cdot  \langle \vec{\sigma} \rangle_2}{6 m_1 m_2} \right] \delta^{(3)}\left(\vec{r} \right) \right\}
\label{V-V_s=1/2} 
\end{equation}

\begin{equation}
\Delta V_{PV-PV}^{s=1/2} = + g_{PV}^1 g_{PV}^2 \, \langle \vec{\sigma} \rangle_1 \cdot  \langle \vec{\sigma} \rangle_2 \, \left\{ I(r) +    \left[ \frac{1}{8} \left( \frac{1}{m_1^2} +\frac{1}{m_2^2} \right) - \frac{1}{m^2} \right]  \, \frac{\delta^{(3)}\left(\vec{r} \right)}{3} \right\}
\label{PV-PV_s=1/2} 
\end{equation}

In eq.\eqref{S-S_s=1/2} the correction term modifies the full potential by eliminating a momentum-independent monopole-monopole term $\sim m^2 \left( \frac{1}{m_1^2} +\frac{1}{m_2^2} \right) \frac{e^{-mr}}{4\pi r}$. A similar modification, but momentum dependent, takes place in eq.\eqref{V-V_s=1/2}. For the other potentials the modifications involve spin-spin interactions, mostly coupled to contact terms.


\subsection{Potentials (spin-1)} \label{sec_spin_1}

\indent

The case of spin-1 is simpler relative to the one from spin-1/2, since the normalization factors are energy independent. The corrections to the affected potentials (vector and tensor representations) are presented below.

\subsubsection{Vector representation}

\indent 

We initiate with the corrections for the vector representation. The corrections to the eqs.(28), (30), (47) and (48) in \cite{grupo} are respectively


\begin{equation}
\Delta V_{S-S}^{s=1} =  \frac{ g_{S}^1 g_{S}^2}{2} \frac{\delta_1 \delta_2}{m_1 m_2} \, \left\{ I(r) - \frac{1}{48} \left( \frac{1}{m_1^2} +\frac{1}{m_2^2} \right) \, \delta^{(3)}\left(\vec{r} \right) \right\}  
\label{S-S_s=1_vet} \end{equation}

\begin{equation}
\Delta V_{PS-PS}^{s=1} = - g_{PS}^1 g_{PS}^2 \, \frac{\langle \vec{S} \rangle_1 \cdot  \langle \vec{S} \rangle_2}{3} \, \delta^{(3)}\left(\vec{r} \right)
\label{PS-PS_s=1_vet} \end{equation}

\begin{equation}
\Delta V_{V-V}^{s=1} = - e_1 e_2   \left\{ 2 \delta_1 \delta_2 \, I(r) + \left[ \frac{2}{3} \frac{\langle \vec{S} \rangle_1 \cdot  \langle \vec{S} \rangle_2}{ m_1 m_2} \, + \frac{\delta_1 \delta_2}{6} \left( \frac{1}{m_1^2} +\frac{1}{m_2^2} \right) \right] \delta^{(3)}\left(\vec{r} \right) \right\},
\label{V-V_s=1_vet} 
\end{equation}

\begin{equation}
\Delta V_{PV-PV}^{s=1} =  g_{PV}^1 g_{PV}^2 \, \frac{\langle \vec{S} \rangle_1 \cdot  \langle \vec{S} \rangle_2}{12 m_1 m_2} \, \delta^{(3)}\left(\vec{r} \right)
\label{PV-PV_s=1_vet} \end{equation}
whereby, in eq.\eqref{V-V_s=1_vet}, $I(r)$ cancels another term in eq.(47) from \cite{grupo}.

\subsubsection{Tensor representation}

\indent 
 
Now we present the corrections for the tensor representation. The correction to the PS-PS potential in this representation is $\Delta V_{PS-PS}^{s=1}/4m_1^2m_2^2$ (cf. eq.\eqref{PS-PS_s=1_vet}). Also, the correction to the PV-PV potential in this representation is the same as in eq.\eqref{PV-PV_s=1_vet}.

In \cite{grupo} eqs.(65) receive the respective corrections:

\begin{equation}
\Delta V_{S-S}^{s=1} =  \frac{g_{S}^1 g_{S}^2}{2} \frac{ \delta_1 \delta_2  }{m_1 m_2} \, \left\{ I(r) - \frac{5}{48} \left( \frac{1}{m_1^2} +\frac{1}{m_2^2} \right) \, \delta^{(3)}\left(\vec{r} \right) \right\}  
\label{S-S_s=1_tens} \end{equation}

\begin{equation}
\Delta V_{V-V}^{s=1} =  - e_1 e_2   \left\{ 2 \delta_1 \delta_2 \, I(r) + \left[ \frac{2}{3} \frac{\langle \vec{S} \rangle_1 \cdot  \langle \vec{S} \rangle_2}{ m_1 m_2} \, + \frac{\delta_1 \delta_2}{3} \left( \frac{1}{m_1^2} +\frac{1}{m_2^2} \right) \right] \delta^{(3)}\left(\vec{r} \right) \right\}.
\label{V-V_s=1_tens} 
\end{equation}

We note that the correction in eq.\eqref{V-V_s=1_tens} excludes
one momentum-dependent term, associated with $ \vec{p}^{\,2} \left( \frac{1}{m_1^2} +\frac{1}{m_2^2} \right) \frac{e^{-mr}}{4\pi r}$, in the V-V potential in eq.(65) from \cite{grupo}.


\section{Concluding remarks} \label{sec_conc}

\indent

In this short note we presented the corrections to some of the results given in Ref.~\cite{grupo}. The corrections introduced affect mostly the  monopole-monopole sectors (e.g., eqs.\eqref{S-S_s=1/2}, \eqref{S-S_s=1_vet} and \eqref{S-S_s=1_tens}) and introduce new spin-dependent contact terms.

The general conclusions presented in \cite{grupo} remain unaltered: in general, the interparticle potentials associated to non-relativistic spin-1/2 and spin-1 sources present similarities, specially in the spin-spin (dipole-dipole) sectors of the $S-S$, $S-PS$, $PS-PS$ and $V-V$ interactions. The most important modifications are due to the presence of contact terms, but these do not play a role in macroscopic interactions.









